\documentclass{nature}

\usepackage{float}
\usepackage{url}

\usepackage{graphicx}
\usepackage{amssymb}
\usepackage{amsmath}

\title{Localized Stress Fluctuations Drive Shear Thickening in Dense Suspensions}

\author{Vikram Rathee$^{1}$, Daniel L. Blair$^1$ \& Jeffrey S. Urbach$^1$}

\begin{document}

\maketitle

\begin{affiliations}
\item Dept. of Physics and Institute for Soft Matter Synthesis and Metrology, Georgetown University, Washington, DC.
\newline{Email: vi21@georgetown.edu; dlb76@georgetown.edu; urbachj@georgetown.edu.}
\end{affiliations}



The mechanical response of solid particles dispersed in a Newtonian
fluid exhibits a wide range of nonlinear phenomena including a
dramatic increase in the viscosity~\cite{Barnes, Melrose_1,
  Wagner_1} with increasing stress. If the volume fraction of the solid phase is moderately
high, the suspension will undergo continuous shear thickening (CST),
where the suspension viscosity increases  smoothly
with applied shear stress; at still higher volume fractions the
suspension can display discontinuous shear thickening (DST), where the
viscosity changes abruptly over several orders of magnitude upon
increasing applied stress. Proposed
 models to explain this phenomenon are based in two distinct
types of particle interactions, hydrodynamic\cite{Melrose_1, Foss, wagner_05} and
frictional~\cite{Seto, Isa, Cates_1, Poon_1, Jaeger_1}. In both cases, the
increase in the bulk viscosity is  attributed to some form
of localized clustering~\cite{Itai_1, Royer}. However, the physical properties and dynamical behavior of these
heterogeneities remains unclear. Here we show that continuous shear thickening
originates from dynamic localized well defined regions of particles with
a high viscosity that increases rapidly with
concentration. Furthermore, we find that the spatial extent of these regions
is largely determined by the distance between the shearing
surfaces. Our results demonstrate that continuous shear thickening
arises from increasingly frequent localized discontinuous transitions
between coexisting low and high viscosity Newtonian fluid phases. Our
results provide a critical physical link between the microscopic
dynamical processes that determine particle interactions and bulk rheological response of
shear thickened fluids.

Recent experimental work
suggests that both frictional and hydrodynamic lubrication   forces play a
significant role in CST \cite{Itai_1, Royer}.  Wyart and Cates (W-C)
have introduced a phenomenological model in which the shear thickening
arises from a transition from primarily hydrodynamic interactions when
the applied stress is substantially below a critical stress, $\sigma
\ll \sigma ^*$, to primarily frictional interactions when $\sigma \gg
\sigma ^*$.
While the W-C model and its extensions can reproduce the measured
average rheological behavior \cite{Cates_1,Royer}, it does not account
for observed temporal fluctuations in bulk viscosity or shear rate
\cite{Lootens, Claus}, or local density differences observed in
magnetic resonance studies of a corn-starch suspension \cite{Ovarlez}.
Similarly, numerical simulations have revealed stress fluctuations at
the particle level and at larger scales \cite{Claus}.  These
results indicate that the spatiotemporal dynamics of shear thickening
includes complexity that cannot be captured by mean field models like
W-C.


Here we describe the result of directly measuring the local
stresses transmitted to one surface of a sheared suspension using {\em
  Boundary Stress Microscopy} (BSM), a technique we developed that
applies traction force microscopy to rheological
experiments\cite{Rich}. To perform BSM measurements, we replace the
bottom rheometer plate with a glass slide coated with a thin, uniform
elastomer of known elastic modulus with a sparse coating of bound
fluorescent  microspheres. We use the measured displacements of the
microspheres to determine the stresses at the boundary with high spatial
and temporal resolution (see Methods). We apply BSM to a suspension of silica
colloidal particles with radius $a = 0.48\,\mu$m 
suspended in an
index matched mixture of glycerol and water. Rheology is performed
using a 25mm cone-plate geometry to ensure a uniform average shear
rate, with a fixed, flat bottom surface and an angled cone that defines the top surface of the sample and rotates in response to the torque applied by the rheometer (Supplementary  Fig. 1a).

The rheological response of these suspensions to increasing stress is
similar to the typical shear thickening behavior previously
reported~\cite{Barnes}.  Specifically, we observe that the measured
viscosity $\eta$ as a function of the applied shear stress $\sigma$,
is strongly non-Newtonian for concentrations $\phi =V_p/V > 0.3$, 
 where $V_p$ and
$V$ are the total particle volume and the system volume respectively (Fig. 1a).
 When $\phi = 0.35$, we observe a
moderate shear thinning for $\sigma < 10$ Pa, followed by a Newtonian
plateau and finally shear thickening, indicated by a small increase in
$\eta$ above an onset or critical shear stress $\sigma_{c}$
(Fig. 1a, open circles). For concentrations of $\phi
\ge 0.5$, strong thinning is followed by an increase in the plateau
viscosity and above $\sigma_{c}$, $\eta$ scales with an exponent $\beta$ ($\eta \propto \sigma ^ \beta$) that increases
 with $\phi$
(Fig. 1a). As previously reported, $\sigma_{c}$ does
not show significant concentration dependence \cite{Wagner_2, Royer}.
For this system we observe discontinuous thickening $\beta=1$ at
$\phi= 0.58$; here we focus on concentrations below this value.

While the temporally averaged response provides a bulk picture of CST,
the time resolved rheology reveals substantial fluctuations.  Figure 
1b shows the strain rate reported by the rheometer
as a function of time for suspensions at $\phi = 0.56$ with a constant
applied stress for 180 seconds.  The fluctuations represent variations
in the shear rate as the rheometer adjusts the rotation rate of the
cone that defines the upper boundary of the suspension in order to
maintain a constant total stress.  For stresses below the onset of
shear thickening, $\sigma_c \sim 50$ Pa, fluctuations are absent (not
shown). We observe that as the stress is increased above $\sigma_c$
the rapid increase in the viscosity is associated with a marked
increased in the magnitude of the fluctuations
(Fig. 1b).  Below $\phi < 0.54$, we do not see any
fluctuations in $\dot{\gamma}$ (Supplementary Fig. 2), although $\eta$
does increase substantially above $\sigma_c$.

In addition to measuring the system-averaged viscosity versus time
under constant stress, we measure the spatially resolved boundary
stresses at specific locations in the rheometer using BSM.  Below
$\sigma_c$ the displacements of the elastic substrate, and therefore
the calculated boundary stresses, are spatially and temporally uniform
(Supplementary Fig. 1b).  However, above $\sigma_c$ and for
concentrations $\phi \ge 0.52$, we observe the appearance of localized
surface displacements that are much higher than the average displacement
(Supplementary Fig. 1c).   

Figure
1c  shows an example of the spatial map the component of the boundary stress in the velocity direction calculated from the measured displacement fields.  The field of view is $890\times 890 \ \mu$m and the
regions of high stresses are large compared to the particle
size. Moreover, the total area imaged represents $0.16\%$ of the
total surface area of the system, indicating that the spatial scale of the stress variation is 
much smaller
than the scale of the lateral system size (25 mm).
Images at higher spatial
resolution do not reveal any any significant stress variations on
smaller length scales, down to our instrumental resolution of $\sim 2
\mu$m \cite{Rich}.

\begin{figure}
\includegraphics[scale=0.3]{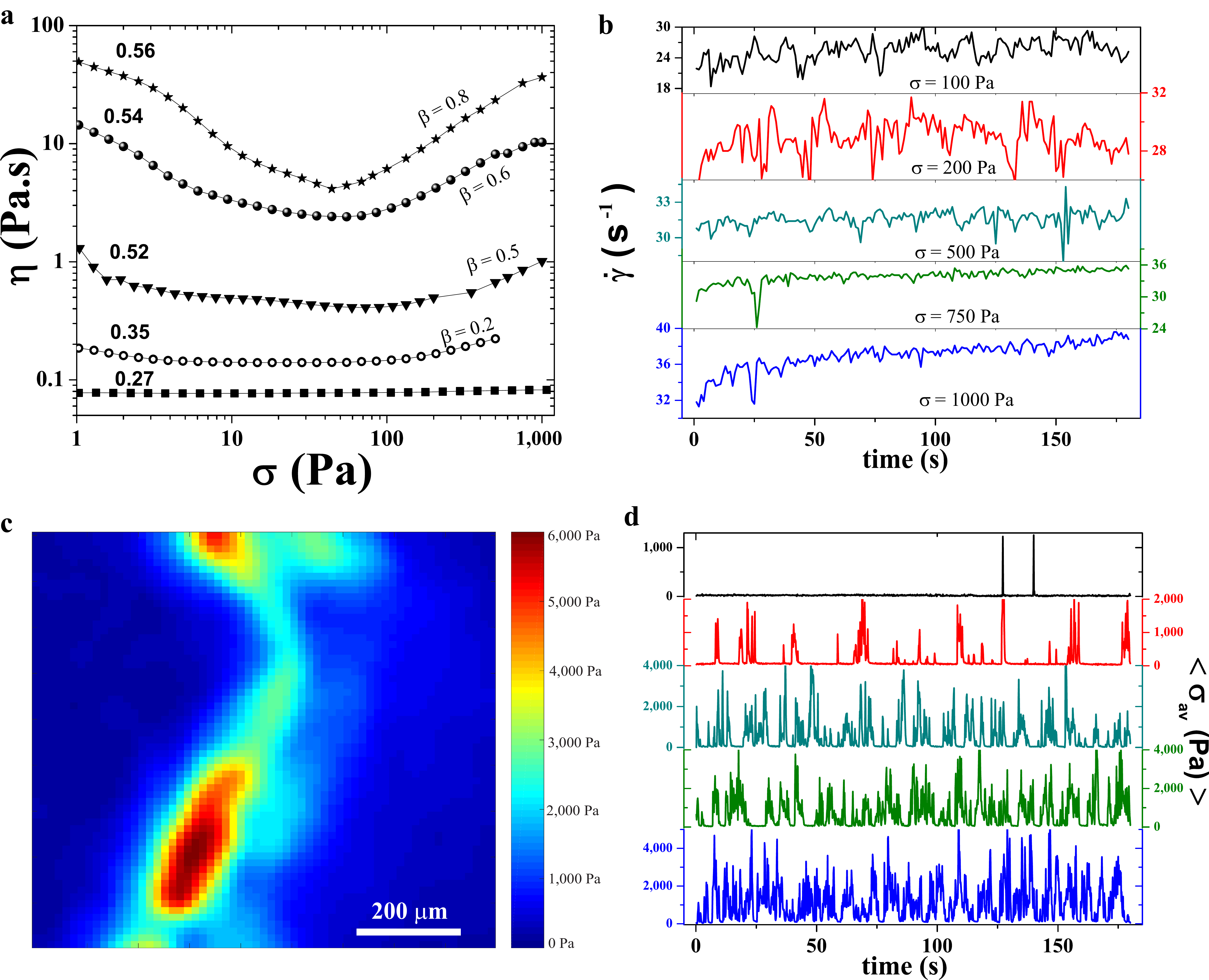}
\caption{figure}{\textbf{Figure 1 $\mid$ Rheological and boundary stress microscopy (BSM)
    measurements of sheared suspension.}  (a) Viscosity vs. stress
  flow curves for suspensions of different concentrations $\phi$, with
  the approximate shear thickening exponent $\beta$; (b) Time
  dependence of the viscosity while shearing at constant applied
  stress ($\phi$ = 0.56) (c) Snapshot of the component of the  boundary stress in the velocity direction
  captured at $\phi$ = 0.56 and $\sigma$ = 1000 Pa (d) Time series of
  average stress per frame from BSM ($\phi$ = 0.56).
\label{fig:bulk_rheo} }
\end{figure}

The regions of high stress at the boundary appear with increasing
frequency as the applied stress in increased above $\sigma_c$. This
can be clearly seen in a time series of the average stress in each
image (see Fig. 1d).   At applied stresses just
above $\sigma_c$, most images do not exhibit high stress regions,
and thus have a small average stress. High stress fluctuations are
clearly separated from the smooth background and appear intermittently
separated by large quiescent periods.  Even the relatively short
regions of low stress evident in the data at $\sigma=$ 500 Pa indicate a large
number of strain units without high stress fluctuations, as can be
seen by plotting the average stress versus strain rather than time
(Supplementary Fig. 3).  Significant local stress fluctuations are
also observed at $\phi$ = 0.52 for applied stresses above $\sigma_{c}$
(Supplementary Fig. 4), although they are both less frequent and
smaller in magnitude.

The high stress regions appear and evolve stochastically, with an
overall average translation in the velocity direction (see
Supplementary Videos 1-6).  A particularly clear example of that motion is
shown as a series of snapshots in Figure 2a, where a high
stress event that is extended in the vorticity direction moves across
the field of view over the course of 3 frames (2/7 s).  The average size of the high stress regions can be 
assessed from the 2-D spatial
autocorrelation of the component of the boundary stress in the
velocity direction.  Figure  2b compares the spatial correlation along the velocity direction from BSM measurements taken half way between the center of the
rheometer and the edge, where the gap between the elastic layer and
the cone is $h=100 \, \mu$m, with those taken close to the outer edge of the system, where $h=200 \, \mu$m.  
 The autocorrelation curves do not appear to
follow a simple functional form, but the comparison suggests that the spatial extent of the high stress regions in the velocity direction is determined by the rheometer gap.  By contrast, the spatial correlation in the vorticity direction does not depend on the gap (Fig.  2b, inset).

The average
motion of the high stress regions can be quantified by cross-correlating the stress patterns for
different time lags (see Methods).  Figure 2c shows an
example of the temporal cross-correlations along the velocity
direction.  The decrease in the peak height with time lag is a consequence of the
fluctuations as the stress patterns move and disappear erratically,
but the displacement of the peaks is indicative of robust propagation
in the velocity direction.  That propagation velocity scales with the
shear rate, and we find that it is consistently equal to half the
velocity of the upper boundary at that location in the sample, which is the velocity of the suspension at the
middle of the 100 $\mu$m gap, assuming a symmetric shear profile
(Fig. 2d). These results suggest that the high boundary
stresses reflect high viscosity fluid phases that span the gap of the
rheometer, sheared equally from above and below and at rest in a frame
co-moving with the center of mass of the suspension.  We observe
qualitatively similar behavior where the gap is 200 $\mu$m, but the
cross-correlations do not show clean peaks, likely due to the higher
propagation speed, since the top boundary moves twice as fast that at the
100 $\mu$m gap, and the fact that the size of the high stress regions
is a larger fraction of our field of view.


\begin{figure}
\includegraphics[width=15.6cm]{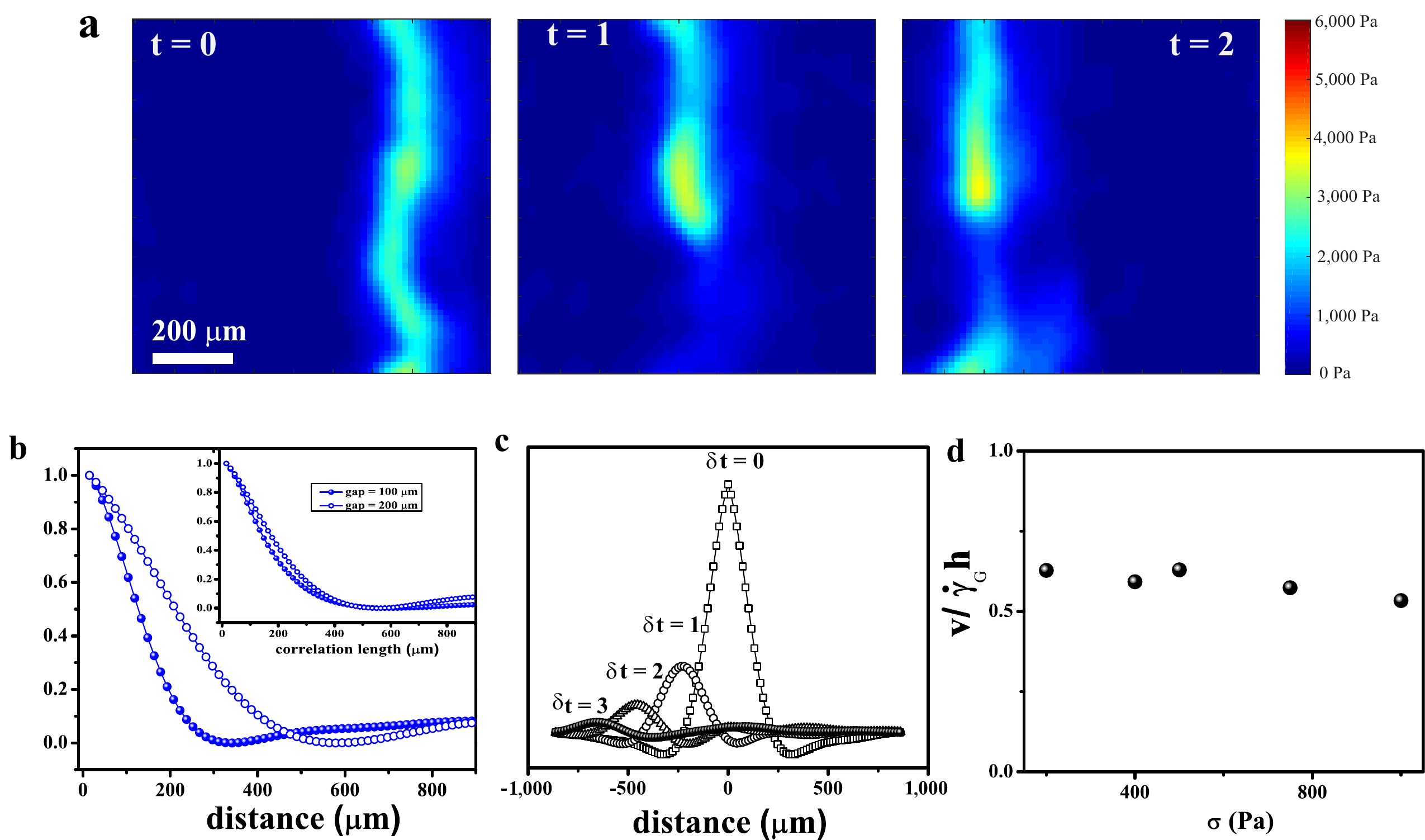}
\caption{figure}{\label{fig:flow} \textbf{Figure 2} $\mid$ (a) Snapshots of the stress field at three consecutive
  time points.  
(b) Line profiles of the normalized two-dimensional autocorrelation of $\sigma_x({\vec{r}})$ along the velocity direction  at different values of the rheometer gap  Inset:  Profiles along the vorticity direction.
  (c) Evolutions of the average temporal
  cross-correlation at different time lags ($\sigma=$ 1000 Pa, $h$ = 100 $\mu$m).  (c) Propagation velocity
  of high stress regions scaled by the velocity of the top boundary ($h$ = 100 $\mu$m).}
\end{figure}

The appearance of localized boundary stresses with magnitudes that are
well above the average stress suggests that the suspension is
spontaneously separating into two distinct and coexisting low and high
viscosity fluid phases during CST, analogous to the low stress and high stress branches of the W-C model, but with dynamics that fluctuate in space and time.  
We can calculate the
viscosity of the low viscosity phase  as
$\eta_L = \sigma_{L}/\dot{\gamma}_{A}$, where $\sigma_{L}$ is the
average stress in all frames that do not have any high stress regions
and $\dot{\gamma}_{A}$ is the average shear rate of the system
reported by the rheometer (see Methods). 
$\eta_L$ shows a modest
increase as the applied stress is increased from 100 Pa to 1000 Pa
(see Fig 3a, filled squares).  We also determine the value of the
high viscosity phase by segmenting the frames showing large stress
heterogeneities into low stress and high stress regions using a
threshold that is well above the stress generated by the low viscosity
fluid (see Methods).  The viscosity calculated from the
average stress in these regions, $\eta_H$ =
$\sigma_{H}$/$\dot{\gamma}_{A}$ is almost two order of magnitude
higher than the low viscosity fluid and nearly constant over the
entire range of applied stress (Fig. 3a, open circles).

The average shear stress exerted by the suspension is the sum of
$\sigma_{L}$ and $\sigma_{H}$ weighted by their respective areas, so
we would expect the average suspension viscosity, $\eta_{A}$, to be
given by
\begin{equation}
\eta_{A} = (1-f)\eta_{L} + f\eta_H,
\label{eq:viscosity}
\end{equation}
where $f$ is fraction of the surface area exhibiting high stresses,
averaged over space and time.  The first term on the right side of
Equation \ref{eq:viscosity} is nearly constant
(Fig. 3b, open squares), as the slight increase in
$\eta_L$ is balanced by an increase in $f$. The second term,
representing the contribution of the high viscosity phase and shown by
filled circles in Fig. 3b, increases dramatically
and clearly captures most of the shear thickening behavior. The
average viscosity, $\eta_{A}$, calculated from the local stress
measurement at h = 200 $\mu$m (Fig. 3b, open
triangles), closely matches that measured by the rheometer
(Fig. 3b, filled triangles) except at the lowest
applied stress value where the high boundary stress regions occur
extremely infrequently, making a statistically meaningful calculation
of the average contribution of the high viscosity phase difficult.
These data show that the smooth increase in viscosity measured by bulk
rheology in CST is in fact the result of increasingly frequent
localized discontinuous shear thickening events. Both $\eta_L$ and
$\eta_H$ for the two gaps are similar at all stresses, indicating that
the intrinsic properties of the two distinct fluid phases are not
dependent on $h$ (Fig. 3c).  However, $\eta_H$
increases dramatically with increasing $\phi$, while remaining roughly
independent of $\sigma$ (Fig. 3d), suggesting that
the high viscosity fluid is approaching something like a jammed state
at $\phi \sim 0.58$, similar to the frictional branch of the W-C
model.

\begin{figure}
\includegraphics[width=15.6cm]{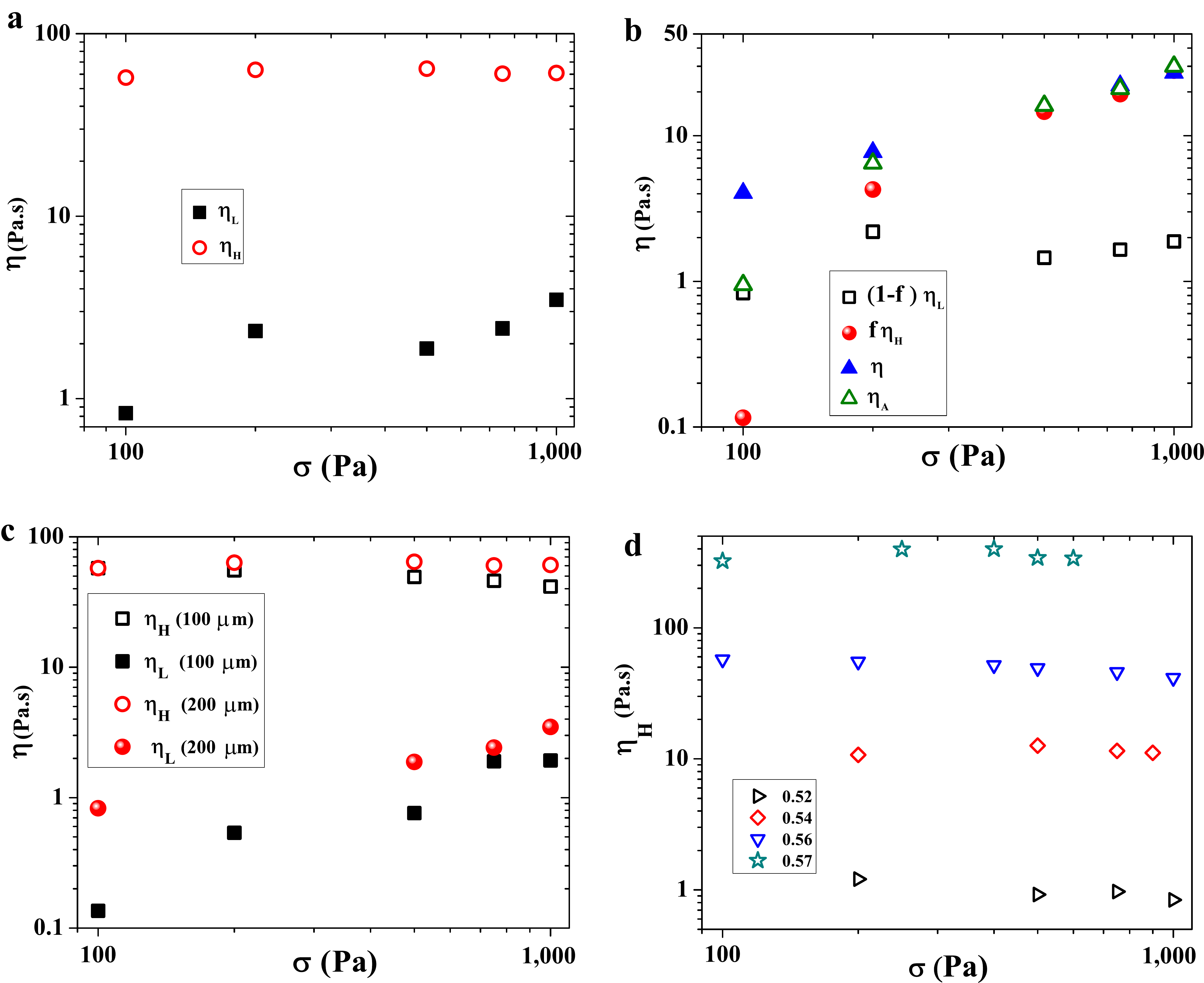}
\caption{figure}{\label{fig:viscosity}\textbf{Figure 3 $\mid$ Viscosities of two distinct fluid phases.}  (a) Viscosity of low viscosity
  fluid ($\eta_{L}$, closed symbols) calculated for frames when no
  high stress regions are present, and viscosity of high stress
  regions (open circles) calculated from the average stress within
  those regions ($\phi=0.56$, 200 $\mu$m gap). (b) Calculated contribution to the average viscosity
  from the low viscosity ($(1-f)\eta_L$) and high viscosity
  ($f\eta_H$) fluids, and the total estimated average viscosity
  $\eta_A$, compared with the bulk viscosity measurement (closed
  triangle). ($\phi=0.56$, 200 $\mu$m gap) (c) Comparison of the viscosities of the two fluid phases at different values of the rheometer gap ($\phi=0.56$). (d) Viscosity of the high viscosity phase at different concentrations (100 $\mu$m gap).}
\end{figure}

Dramatically different behavior is observed in a suspension of $\phi =
0.57$, closer to the discontinuous shear thickening regime.  The
boundary stress is highly heterogeneous, but in many cases the high
stress regions do not propagate, and instead appear stationary
relative to the bottom surface. Moreover, we find that the particles
just above the bottom plate intermittently become motionless,
indicating that a localized portion of the suspension has formed a
fully jammed, solid-like phase (Supplementary Video 7).  By contrast,
particles in the suspension at $\phi = 0.56$ show continual relative
motion, with no periods of local rigidity, indicating that the
high viscosity phase remains fluid-like (Supplementary Video 8).
While not the primary focus of the current study, the behavior at
$\phi = 0.57$ is discussed in more detail in the Supplementary
Material.

Based on our boundary stress microscopy measurements, the suspension
behavior can be divided into three different regimes (i) $0.3 < \phi
< 0.52$, (ii) $0.52 \leq \phi < 0.57$ and (iii) $\phi \geq 0.57 $. In
regime (i) the boundary stresses are uniform and the bulk shear
thickening exponent $\beta <$ 0.5. In regime (ii), propagating regions
of high stress appear with a frequency and intensity that increases
with applied stress and concentration, and that increase accounts the
observed shear thickening. Regime (iii) is characterized by the
appearance of very high non-propagating boundary stresses, indicative
of fully jammed regions.

The behavior in regime (ii), where regions of the suspension
transition abruptly to a high viscosity phase, is reminiscent of the
bifurcation between low viscosity and high viscosity branches
described in the W-C model, representing states with different
fractions of frictional contacts created when the stress between
particles exceeds the critical stress necessary to overcome the
repulsive forces that stabilize the suspension.  Because the value of $\phi$ at which the viscosity diverges
depends on the
fraction of frictional contacts, an instability arises in certain
regimes due to positive feedback between the increased local stress
and viscosity resulting from the added frictional contacts
between particles.  Our results are consistent with a model where the high stress state appears intermittently in localized clusters with a size determined by the
rheometer gap and with a viscosity that is nearly Newtonian and diverges at $\phi \sim
0.58$.  In the W-C model the instability is present only over a narrow
range of concentrations very close to $\phi=0.58$, and the continuous
shear thickening regime is characterized by a smooth increase in the
fraction of frictional contacts as the applied stress increases.  Our
results imply that   the fraction
of frictional contacts is spatially heterogenous and dynamically
fluctuating throughout the shear thickening regime, suggesting that a refinement of the theoretical models is necessary to correctly capture the range over which discontinuous viscosity transitions can occur.   Moreover,
spatial and temporal variations in the relevant structural and
mechanical fields must be considered, as we find that continuous shear
thickened colloidal suspensions are actually comprised of a spatially
heterogeneous and rapidly evolving regions of distinct high and low
viscosity fluids.

\noindent\textbf{Methods}
Elastic substrates were deposited on glass cover slides of diameter 43
mm by spin coating Polydimethylsiloxane (PDMS) (Sylgard 184®, Dow
Corning) and a curing agent. The PDMS and curing agent were mixed in a
ratio of 35:1 and degassed until there were no visible air
bubbles. Films of thickness 35 $\mu$m were created using spin coating
on cover slides cleaned thoroughly by plasma cleaning and rinsing with
ethanol and de-ionized water.  After deposition of PDMS, the slides
were cured at 85 $^{o}$C for 45 minutes. After curing, the
PDMS was functionalised with 3-aminopropyl triethoxysilane (Fisher
Sci.) using vapor deposition for approximately 40 minutes. For
imaging, carboxylate-modified fluorescent spherical beads of size 1.04
$\mu$m (0.96 $\mu$m) with excitation/emission at 480/520 (660/680)
were attached to the PDMS surface. Before attaching the beads to
functionalised PDMS, the beads were suspended in a solution containing
3:8mg/mL sodium tetraborate, 5mg/mL boric acid and 0:1mg/mL 1-ethyl-
3-(3-dimethylaminopropyl)carbodiimide (EDC) (SigmaAldrich). The
concentration of beads used was 0.006 $\%$ solids.

Suspensions were formulated with silica spheres of radius a = 0.48
$\mu$m (Bangs Laboratories, Inc.) suspended in a glycerol water
mixture (0.8 glycerol mass fraction). The volume fraction is
calculated from the mass determined by drying the colloids in an oven at
100 $^{o}$C.

Rheological measurements were performed on a stress controlled
rheometer (Anton Paar MCR 301) mounted on an inverted confocal (Leica
SP5) microscope \cite{Dutta} using a 25 mm diameter cone with a $1^o$
angle. A 10X objective was used for imaging and produced a $0.89
\times 0.89$ mm$^{2}$ field of view. Images were collected at a rate
of 7 frames s$^{-1}$.  Deformation fields were determined with
particle image velocimetry (PIV) in imageJ \cite{Tseng}.  For some
experiments, fluorescent beads with excitation/emission at 480/520
were added to the silica suspension as tracer particles in ratio of
1:100. In that case, we have attached beads with a different
excitation/emission spectra 660/680 on the PDMS and utlized 480/520
tracer beads and imaged the system with two separate laser lines. The
surface stresses at the interface are calculated using an extended
traction force technique and codes given in \cite{Style}. In all cases
the reported stresses are the component of the boundary stress in the
velocity ($x$)-direction.  2D spatial autocorrelations $<\delta\sigma_x(\vec{r},t)\delta\sigma_x(\vec{r}+\vec{\delta r},t)>$, averaged over $\vec{r}$ and $t$, are calculated using the Matlab function  xcorr2.  1D temporal cross-correlations, $<\delta\sigma_x(x,t)\delta\sigma_x(x+dx,t+\delta t)>$, are calculated similarly.
The average stress in the high stress regions is calculated as 
\begin{equation}
\sigma_{H}  = \langle\sigma_{x}(\vec{r_{i}})\rangle 
\end{equation}   
where the average is taken only over the positions $\vec{r_{i}}$  that satisfy the condition that $\sigma_{x}(\vec{r_{i}}) >$ 500Pa, as indicated by the circles in  Supplementary Fig. 5(d). 
Supplementary Figures 5(a-e) shows how the regions identified as high stress evolve as the threshold is changed.
The absolute values of $\sigma_{H}$ are only weakly dependent on the threshold for the value used here. 
 The average stress produced by the low stress phase is calculated by averaging $\sigma_{x}(\vec{r_{i}})$ from those images where the stress value at every position satisfies the condition $\sigma_{x}(\vec{r_{i}}) < $ 500Pa.


\noindent\textbf{Acknowledgements:} This work was supported by AFOSR
grant FA9550-14-1-0171. J. S. U. is supported in part by the
Georgetown Interdisciplinary Chair in Science Fund D.L.B is supported
in part by the Templeton Foundation grant 57392. We thank Emanuela Del
Gado, David Egolf, Peter Olmsted and John Royer for helpful
discussions and Eleni Degaga and Jim Wu for their contributions to
protocol development.


\newpage
\noindent\textbf{\Large{Supplementary Information}}\\

\noindent\textbf{Behavior Close to Discontinuous Shear Thickening ($\phi$ = 0.57)}:
When the suspension is subjected to shear at $\phi$ = 0.57, close to discontinuous shear thickening concentration  ($\phi=0.58$ for this system), we observe a  slightly higher thickening exponent $\beta$ $\sim$ 0.9 (Supplementary Fig. 6a). The temporal behavior of $\eta$ (Supplementary Fig. 6b) is  similar to that seen at lower concentrations (Fig. 1b).  The localized boundary stresses, however, show markedly different behavior, with the regions of high stress appearing stationary rather than propagating in velocity direction. The stresses within these regions are approximately an order of magnitude of larger than applied stress (Supplementary Fig. 6c). The average local viscosity calculated from BSM matches fairly well with the bulk viscosity (Supplementary Fig. 6d).  We visualized the motion of the suspension in this regime with tracer particles with an excitation/emission different from the fiduciary beads on the PDMS surface. We find that  the tracer particles intermittently  become motionless, indicating that a localized portion in the suspension has jammed into a fully solid (non-sheared) state.  These localized motionless regions often extend almost to the top plate. Supplementary Video 7 shows a representative movie of the suspension 50 $\mu$m above the bottom boundary (at a radius where the gap is 100 $\mu$m).  The motionless regions, since they are finite in extent, are  accompanied by complex flow fields in the surrounding suspension. Interestingly, we observe a large degree of slip at the top plate while jammed region remain static, suggesting a fracture plane is formed at the top of the solid phase.

The existence of a fully jammed solid-like phase is reminiscent of the behavior of W-C model, where at sufficiently high concentrations the strain rate goes to zero above a critical stress\cite{Cates_supp}.  The fact that the boundary stresses associated with that solid phase are motionless indicates that the thin fluid layer that, at lower concentrations, exists between the suspension and the PDMS layer and supports a small boundary slip, is eliminated, presumably due to the dilatancy of the fully jammed suspension. 
This behavior is consistent with the protrusion observed in colloidal as well granular suspensions on the free surface near the outer open boundary during shear thickening\cite{Jaeger_supp}.  These observations highlight the importance of the boundaries the transition to DST, and to the complexity of the spatiotemporal dynamics in this regime.


\noindent\textbf{Supplementary Video 1:} Time evolution of average stress calculated from boundary stress microscopy (left panel) and dynamic stress map (right panel) at $\phi$ = 0.56, $\sigma$ = 100 Pa and gap = 200 $\mu$m. Total elapsed time is 180 seconds and average $\dot{\gamma}$ = 24.75 s$^{-1}$. Please see the link: 
\newline{\url{https://figshare.com/s/c5615553598830ec1add}}

\noindent\textbf{Supplementary Video 2:} Time evolution of average stress calculated from boundary stress microscopy (left panel) and dynamic stress map (right panel) at $\phi$ = 0.56, $\sigma$ = 500 Pa and gap = 200 $\mu$m. Total elapsed time is 180 seconds and average $\dot{\gamma}$ = 31.05 s$^{-1}$. Please see the link:
\newline{\url{https://figshare.com/s/57493ea94ce0a8f4d624}}

\noindent\textbf{Supplementary Video 3:} Time evolution of average stress calculated from boundary stress microscopy (left panel) and dynamic stress map (right panel) at $\phi$ = 0.56, $\sigma$ = 1000 Pa and gap = 200 $\mu$m. Total elapsed time is 180 seconds and average $\dot{\gamma}$ = 37 s$^{-1}$. Please see the link:  \newline{\url{https://figshare.com/s/d1aa4bc8ccf7726af384}}

\noindent\textbf{Supplementary Video 4:} Time evolution of average stress calculated from boundary stress microscopy (left panel) and dynamic stress map (right panel) at $\phi$ = 0.56, $\sigma$ = 100 Pa and gap = 100 $\mu$m. Total elapsed time is 180 seconds and average $\dot{\gamma}$ =  16.5 s$^{-1}$. Please see the link:
\newline{\url{https://figshare.com/s/342d689241bb7ae1a37c}}

\noindent\textbf{Supplementary Video 5:} Time evolution of average stress calculated from boundary stress microscopy (left panel) and dynamic stress map (right panel) at $\phi$ = 0.56, $\sigma$ = 500 Pa and gap = 100 $\mu$m. Total elapsed time is 180 seconds and average $\dot{\gamma}$ = 20.7 s$^{-1}$. Please see the link:
\newline{\url{https://figshare.com/s/86434ccbc143cca29d0c}}

\noindent\textbf{Supplementary Video 6:} Time evolution of average stress calculated from boundary stress microscopy (left panel) and dynamic stress map (right panel) at $\phi$ = 0.56, $\sigma$ = 1000 Pa and gap = 100 $\mu$m. Total elapsed time is 180 seconds and average $\dot{\gamma}$ = 24.75 s$^{-1}$.  Please see the link:
\newline{\url{https://figshare.com/s/4cc6eca08ac658f43cf3}}

\noindent\textbf{Supplementary Video 7:} $\sigma$ = 2000 Pa, $\phi$ = 0.57  and suspension is imaged 50 $\mu$m inside. Before 3.56 s and after 93.6 s there is no applied stress. Total elapsed time is 100 seconds and average $\dot{\gamma}$ = 27.3 s$^{-1}$. Please see the link:
\newline{\url{ https://figshare.com/s/24869ada7c18a998568c}}

\noindent\textbf{Supplementary Video 8:} $\sigma$ = 1000 Pa, $\phi$ = 0.56  and suspension is imaged 50 $\mu$m inside. Before 3.86 s and after 94 s there is no applied stress. Total elapsed time is 100 seconds and average $\dot{\gamma}$ = 25.4 s$^{-1}$. Please see the link: 
\newline{\url{https://figshare.com/s/02d50869831df8eb64c0}}

\begin{figure}
\begin{centering}
\includegraphics[scale=0.55]{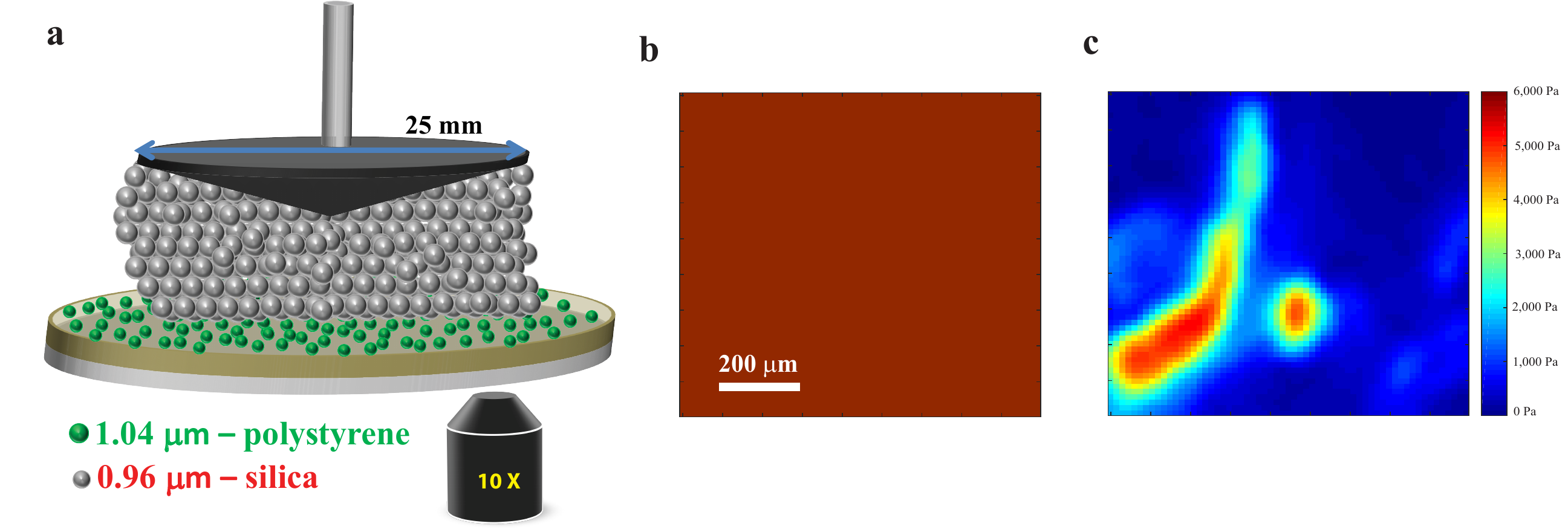}
\end{centering}
\newline{\caption{figure}{ \textbf{Supplementary Figure 1 $\mid$}(a)  Schematic of Boundary Stress Microscopy setup. The bottom cover slide of thickness 100 $\mu$m is shown in silver  and  35 $\mu$m  PDMS layer  is shown in light green. Fluorescent microspheres attached to the PDMS are shown in green. The silica colloids are shown in silver.  The rheometer  cone defines the top surface of the suspension and is attached to a stress controlled rheometer.  The calculated displacement field of the microspheres is shown for a representative uniform (b) and   heterogeneous (c)  case.}}

\end{figure}

\begin{figure}

\includegraphics[scale=0.4]{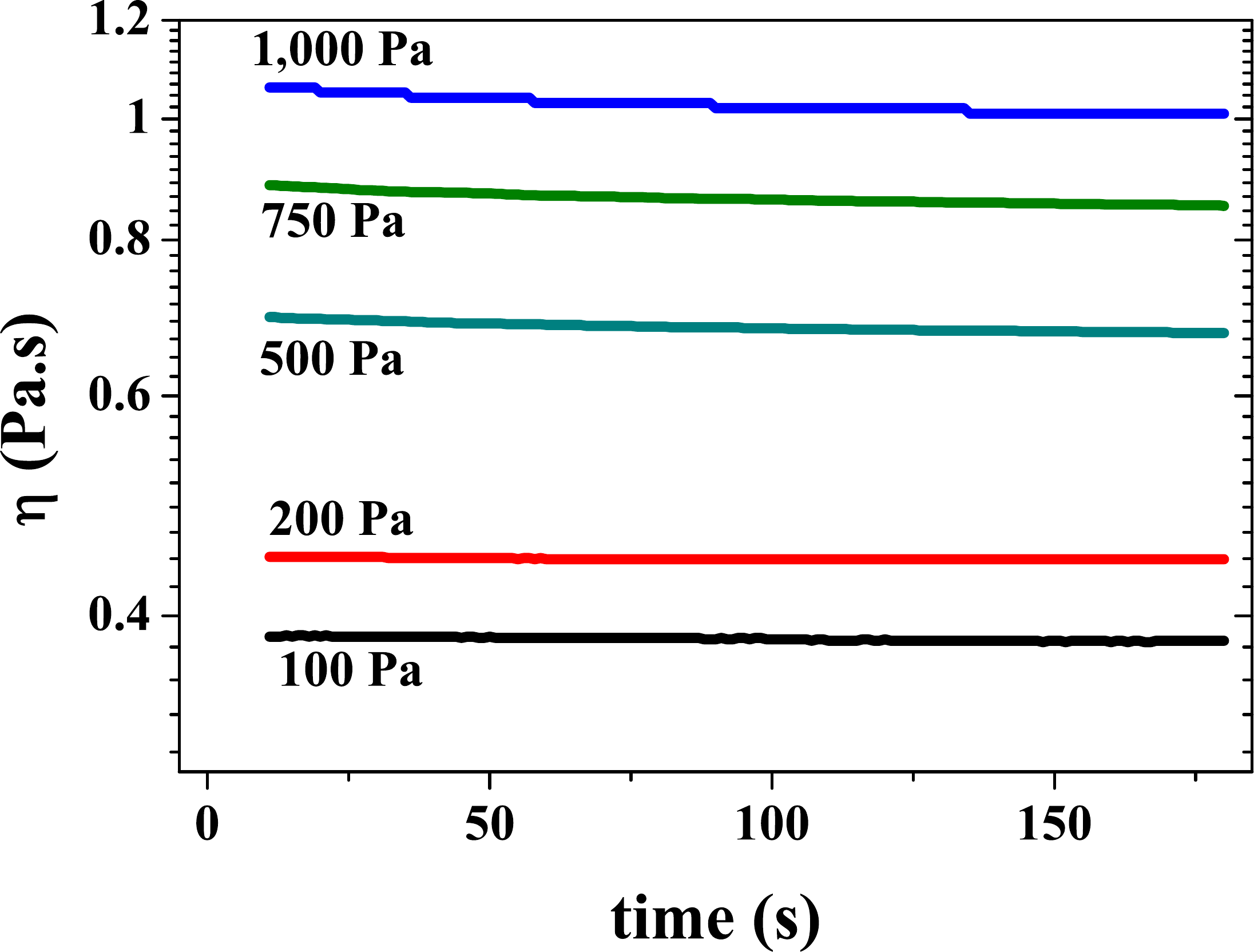}
\newline{\caption{figure}{\textbf{Supplementary Figure 2 $\mid$} Evolution of $\eta$ with time  ($\phi$ = 0.52) at different values of applied stress. }}

\end{figure}

\begin{figure}

\includegraphics[scale=0.5]{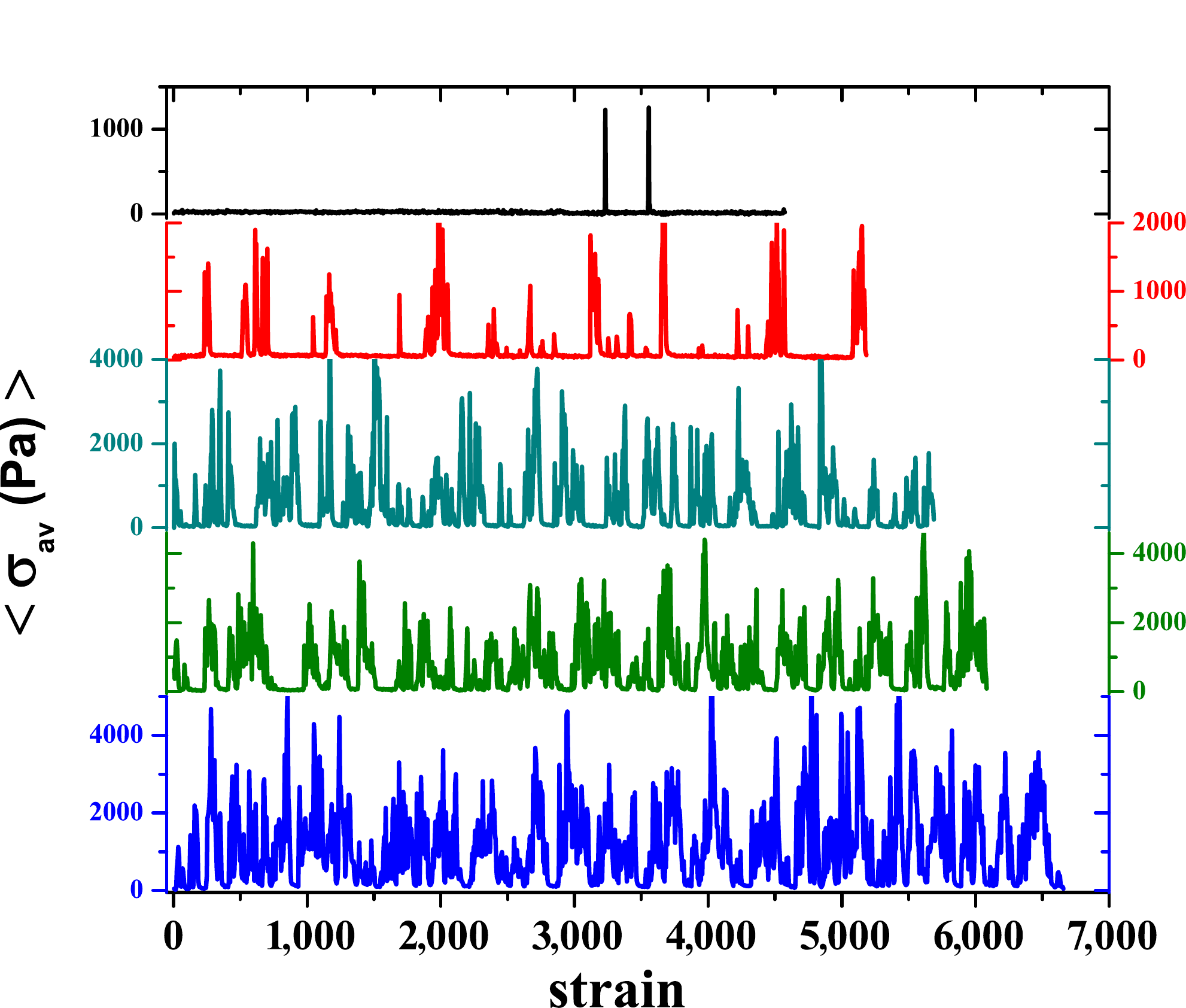}
\newline{\caption{figure}{\textbf{Supplementary Figure 3 $\mid$} Average stress per frame from BSM ($\phi$ = 0.56) plotted as a function of elapsed strain for different values of applied stress  (from top to bottom : 100 Pa, 200 Pa, 500 Pa, 750 Pa and 1000 Pa). }}

\end{figure}

\begin{figure}

\includegraphics[scale=0.45]{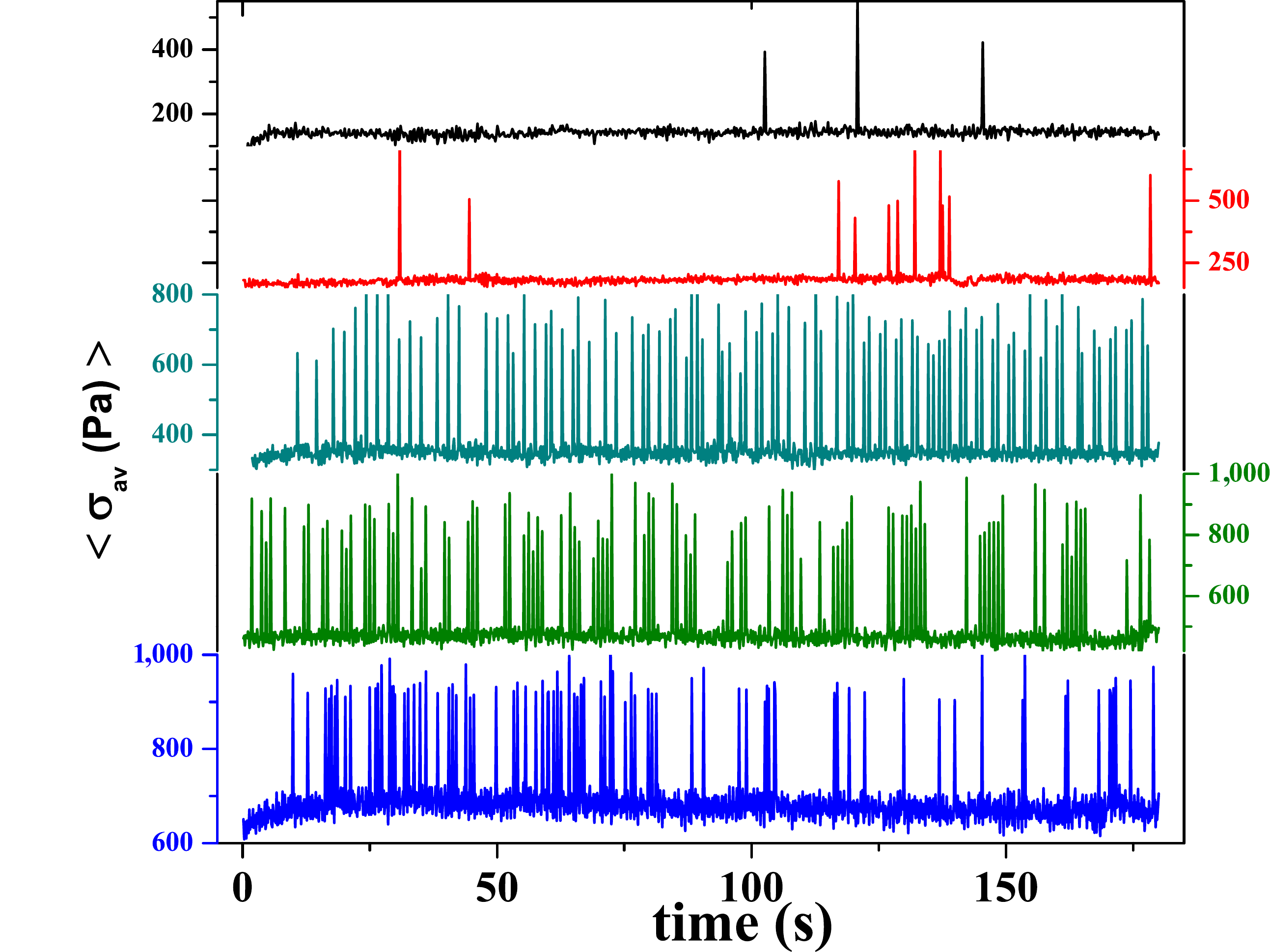}
\newline{\caption{figure}{\textbf{Supplementary Figure 4 $\mid$} Temporal evolution of average stress obtained from boundary stress microscopy for different values of applied stress  (from top to bottom : 100 Pa, 200 Pa, 500 Pa, 750 Pa and 1000 Pa) at $\phi$ = 0.52 and gap = 100 $\mu$m.}}

\end{figure}

\begin{figure}

\includegraphics[scale=0.38]{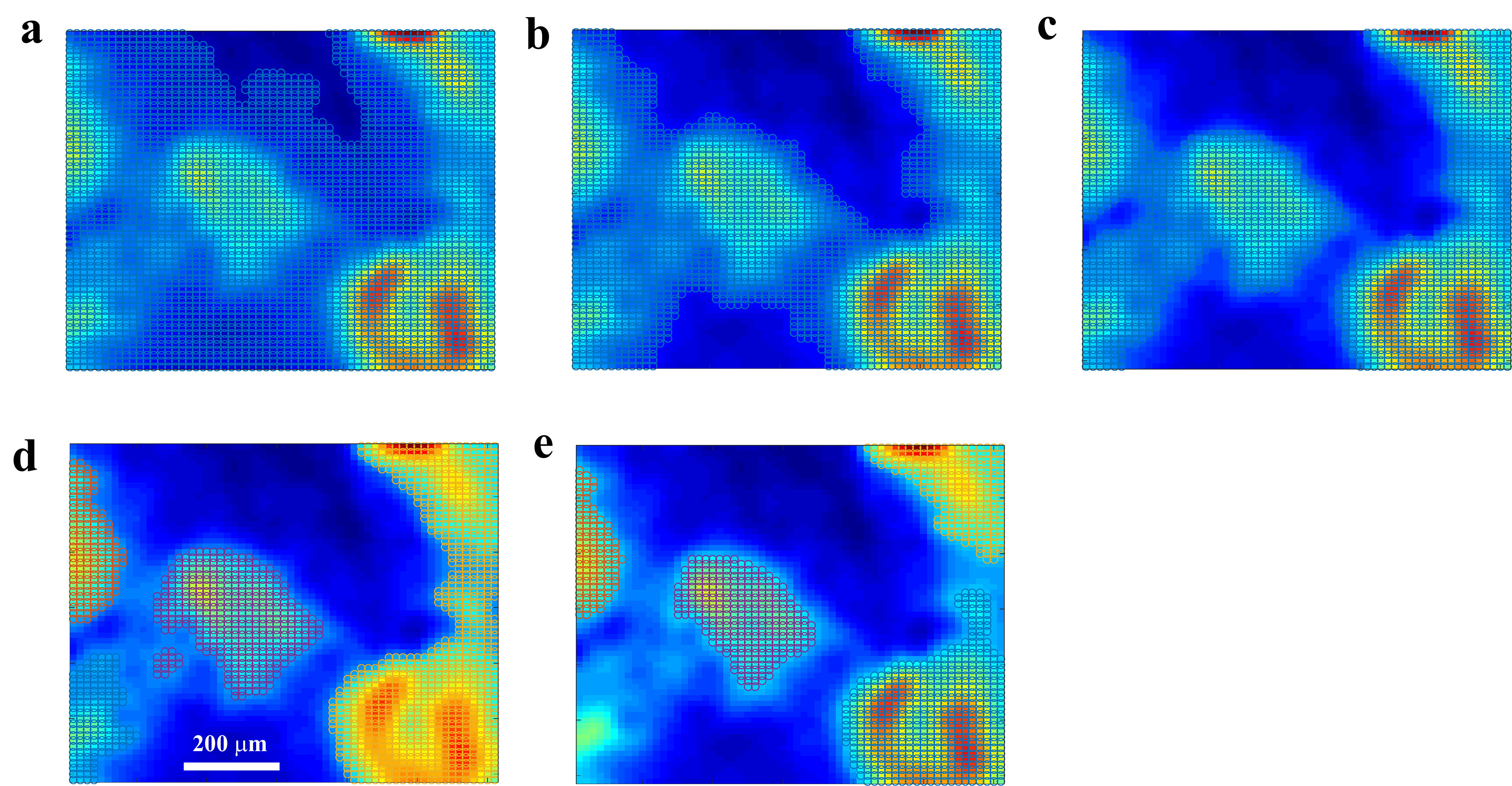}
\newline{\caption{figure}{\textbf{Supplementary Figure 5 $\mid$}Identification of high stress regions for different threshold values.  The circles mark regions above threshold for values 100, 250, 400, 500, and 600 Pa (a- e)  (gap = 100 $\mu$m, $\sigma$ = 1000 Pa).}}

\end{figure}

\begin{figure}

\includegraphics[scale=0.3]{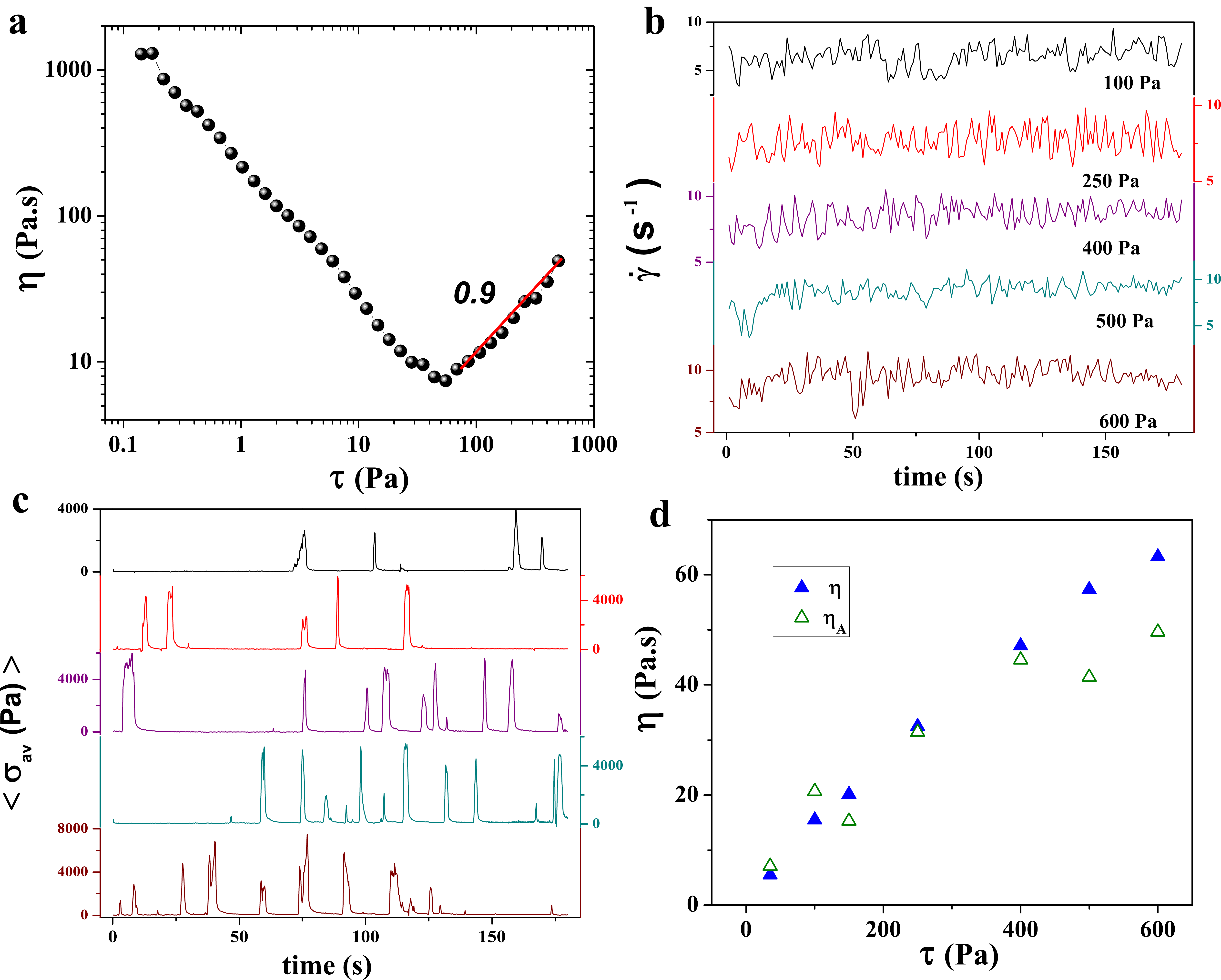}
\newline{\caption{figure}{\textbf{Supplementary Figure 6 $\mid$} a) Viscosity vs stress  flow curve at $\phi$ = 0.57, showing a thickening exponent $\beta$ of 0.9. Temporal evolution of b) shear rate ($\dot{\gamma}$) and c) average stress ($\sigma_{av}$) for applied  $\sigma$ of 100, 250, 400, 500, and 600 Pa, from top to bottom.  d) Bulk viscosity  ($\eta$, closed triangles) and average viscosity ($\eta_{A}$, open triangles) derived using equation (1) of the main text.}}

\end{figure}

\end{document}